\documentclass[12pt]{article}
\usepackage{epsf, amssymb}
\usepackage{epsfig}
\usepackage{hyperref}

\makeatletter
\renewcommand\section{\@startsection {section}{1}{\z@}%
                                   {-3.5ex \@plus -1ex \@minus -.2ex}
                                   {2.3ex \@plus.2ex}%
                                   {\normalfont\large\bfseries}}

\renewcommand\subsection{\@startsection{subsection}{2}{\z@}%
                                     {-3.25ex\@plus -1ex \@minus -.2ex}%
                                     {1.5ex \@plus .2ex}%
                                     {\normalfont\bfseries}}

\makeatother

\newcommand{\bea}{\begin{eqnarray}}
\newcommand{\eea}{\end{eqnarray}}
\newcommand{\be}{\begin{equation}}
\newcommand{\ee}{\end{equation}}

\def\a{\alpha}
\def\b{\beta}

\def\d{\delta}

\def\e{\epsilon}   
\def\f{\phi}               

\def\i{\iota}
\def\im{\mathrm{Im}}
\def\inf{\infty}

\def\l{\lambda}
\def\m{\mu}

\def\na{\nabla}
\def\o{\omega}  
\def\p{\pi}   
\def\pa{\partial}       
\def\re{\mathrm{Re}}                
\def\r{\rho}                                     
\def\s{\sigma}                                   

\def\th{\theta}
\def\til{\tilde}

\def\D{\Delta}

\def\G{\Gamma}

\def\L{\Lambda}
\def\O{\Omega}

\def\S{\Sigma}

\def\cc{{\cal C}}

\def\cn{{\cal N}}
\def\co{{\cal O}}

\def \zz {{\mathbb Z}}
\def \cc {{\mathbb C}}
\def \rr {{\mathbb R}}

\def \yf {\mathsf{y}}
\def \tf {\mathsf{t}}
\def \Lf {\mathsf{L}}
\def \Qf {\mathsf{Q}}
\def \Hf {\mathsf{H}}
\def \hf {\mathsf{h}}

\def \Sf {\mathsf{S}}
\def \xf {\mathsf{x}}

\begin{document}
\begin{titlepage}

\begin{center}

\hfill                  hep-th/0611156

\hfill ITFA-06-40

\vskip 2 cm {\Large \bf More Bubbling Solutions}
\vskip 1.25 cm { 
Miranda C. N. Cheng\footnote{
mcheng@science.uva.nl} }\\
{\vskip 0.5cm  Institute for Theoretical Physics\\ University of Amsterdam\\
Valckenierstraat 65\\
1018 XE, Amsterdam\\
The Netherlands\\}

\end{center}

\vskip 2 cm

\begin{abstract}
\baselineskip=18pt

In this note we construct families of asymptotically flat, smooth, horizonless solutions with a large number of non-trivial two-cycles (bubbles) of \({\cal{N}}=1\) five-dimensional supergravity with an arbitrary number of vector multiplets, which may or may not have the charges of a macroscopic black hole and which contain the known bubbling solutions as a sub-family. We do this by lifting various multi-center BPS states of type IIA compactified on Calabi-Yau three-folds and taking the decompactification (M-theory) limit. We also analyse various properties of these solutions, including the conserved charges, the shape, especially the (absence of) throat and closed timelike curves, and relate them to the various properties of the four-dimensional BPS states. We finish by briefly commenting on their degeneracies and their possible relations to the fuzzball proposal of Mathur {\it{et al}}.

\end{abstract}

\end{titlepage}

\pagestyle{plain}
\baselineskip=19pt

\tableofcontents

\section{Introduction}

The four-dimensional multi-center BPS solutions of type II string theory compactified on a Calabi-Yau three-fold have been derived in  \cite{Behrndt:1997ny,Denef:2000nb,LopesCardoso:2000qm,Denef:2001xn,Bates:2003vx}, and their lift to M-theory was, after the indicative work \cite{Bena:2004tk,Bena:2005ay},  explicitly written down in \cite{Gaiotto:2005xt} (see also \cite{Behrndt:2005he}). Recently, this idea of the 5d lift of 4d multi-center solutions have contributed to the understanding of black ring entropy  \cite{Gaiotto:2005xt,Bena:2005ni,Elvang:2005sa}, the relationship between the Donaldson-Thomas invariants and topological strings \cite{Dijkgraaf:2006um}, and the OSV conjecture \cite{Denef1}. Indeed, with different choices of charges and Calabi-Yau background moduli, one can expect to have a large assortment of BPS solutions to \(\cn=1\) (8 supercharges) five-dimensional supergravity with various different properties by simply lifting various multi-center solutions to five dimensions.

On the other hand, Mathur and collaborators have proposed a picture of black holes different from the conventional one. According to this proposal, the black hole could actually be a coarse-grained description of a large number of smooth, horizonless supergravity solutions (``microstates", ``proto-black holes") which have the same charges as that of a ``real black hole". (see \cite{Lunin:2001jy}, \cite{Mathur:2005zp} and references therein). A question one might then ask is, do there exist some solutions in the zoo of the lifted multi-center solutions which possess this property? If yes, how many of them are there? And how to classify them?

To construct a solution like this via the 4d-5d connection, first of all in order to have the right global feature at spatial infinity (that it should approach \(\rr_t\times \rr^4\) but not \(\rr_t\times \rr^3 \times S^1\)), one would need to take the decompactification limit in which the M-theory circle is infinitely large at spatial infinity. In this limit the five-dimensional description is also the only valid one. Furthermore, for the smooth and horizonless feature we have to restrict ourselves to D6 or/and anti-D6 branes as the centers in 4D. To obtain non-trivial charges we then turn on the world-volume fluxes on these centers. Finally we lift the solutions with these charges and background to five dimensions. In this way we have indeed obtained a large number of asymptotically flat, smooth and horizonless solutions, to five-dimensional supergravity theories with an arbitrary number of vector multiplets, which may have the total charge of that of a black hole. Actually, if we restrict to the STU Calabi-Yau and make a special Ansatz of the K\"ahler moduli, we retrieve the known bubbling solutions of \cite{Berglund:2005vb,Bena:2005va,Bena:2006is}.\footnote{In \cite{Balasubramanian:2006gi} it has been observed that, if one adds a constant term to one of the harmonic functions in the Bena-Warner {\it{et al}} bubbling solutions, which corresponds to de-decompactify the extra dimension, and then reduce it, one would get a 4D multi-center solution. See also \cite{Saxena:2005uk} for a related discussion. } In a recent paper, through a more explicit study of the above-mentioned solutions, Bena, Wang and Warner \cite{Bena:2006kb} have constructed the first smooth horizonless solutions with charges corresponding to a BPS three-charge black hole with a classical horizon. Indeed, to understand this recent development has been the original motivation of the present work.

To be able to have a solution like this in the case of a general Calabi-Yau compactification further heightens the contrast between the picture of a black hole of Mathur {\it{et al}} and the conventional one . Unlike the torus case, a general Calabi-Yau with its complicated topological data is generically the biggest origin of a large black hole entropy \cite{Maldacena:1997de,Vafa:1997gr}. As we have mentioned, to have a horizonless solution lifted from four dimensions forces us to consider only rigid centers, i.e., those without any (classical) internal degrees of freedom associated to them. To reconcile these two pictures therefore seems to be much more challenging in the case of a general Calabi-Yau compactification. The authors of \cite{Balasubramanian:2006gi} have proposed a following picture: while the system is described by a D-brane bound state at weak  string coupling, it expands into a multi-particle system when we turn on the \(g_s\) and is thus described by a multi-centered supergravity solution, and further grows into a five-dimensional system when the string coupling is increased even further. While this picture has been carefully studied and tested in the case with the total charge {\it{not}} corresponding to that of a classical black hole  \cite{Denef:2002ru}, we don't seem to have much evidence to argue the same for the case with black hole total charges. In other words, {\it{a priori}} we don't see the reason why the D-brane bound state must open up into a multi-center configuration instead of staying together and form a black hole in the conventional sense, as \(g_s\) is slowly turned on. To sum up, how one would be  able to reconcile the two pictures of black holes remains mysterious.

The paper is organised as follows: in section two we lay out our notations and review the 4d multi-center BPS solutions and their lift to five dimensions. In section three we construct our bubbling solutions in 3 steps. First we work out the 4d solution in the M-theory \(\Leftrightarrow\) large IIA Calabi-Yau volume limit, and lift it to five dimensions. Secondly we rescale the five-dimensional coordinates to make it commensurable with the five-dimensional Planck units. Finally we put in the charge vectors of D6 and anti D6 with fluxes and arrive at the final form of the bubbling solutions. 

In section four we analyse in full details the various properties of these solutions. A large part of the analysis holds also for generic lifted multi-center solutions in the decompactification limit, and some furthermore also holds for generic values of background moduli. Therefore, along the way we have also derived various properties of all the lifted multi-center solutions; or to say, the properties of various configurations of charged objects in type IIA string theory in the very strong coupling limit. Specifically, in 4.1 we work out the asymptotic metric, read off the five-dimensional conserved charges, including the electric charges of the M-theory C-field, and the two angular momenta \(J_L\) and \(J_R\), for generic centers. In 4.2 we focus on the metric part and first study the condition for the absence of closed timelike curves (CTC's). Here we find a map of diseases: a CTC pathology in 5D corresponds to an imaginary metric pathology in 4D. We also analyse the possibility of having a throat-like (i.e. AdS-looking) metric in some part of the space. We conclude, also independent of the details of how the charges get distributed, that a multi-center configuration with charges not giving any black hole can never have a region like that, at least in the regime where supergravity is to be trusted.  We also check that, for our specific fluxed D6 and anti-D6 composition, the metric is smooth (at worst with an orbifold singularity when there are stacked D6) and horizonless everywhere, and we do this by establishing that the metric approaches that of a(n) (orbifolded) flat \(\rr^4\times\rr_t\) in the vicinity of each center. In 4.3 we briefly discuss the role of the large gauge transformation of the M-theory three-form potential in our setting. We end our paper with discussions about future directions and some more speculative discussions about the degeneracy of ``black holes" or "proto-black holes".

\section{Review: The Lift of Multi-Center Solutions}
\setcounter{equation}{0}

The lift \cite{Gaiotto:2005xt} of the multi-center solution \cite{Bates:2003vx} is the starting point of our construction of the new bubbling solutions. In this section we will briefly review and slightly rewrite them in our conventions. 

Let's first define the Calabi-Yau (X) data: take \(\a_A\), \(A= 1, ..., b_2(X)\) to be a basis of \(H^2(X;\zz)\) and \(\b^A\) its dual basis, i.e. \(\int_X a_A \wedge \b^B = \d^B_A\) .  We also write \(\a_0=1\) and \(\b^0 = \frac{J\wedge J\wedge J}{J^3} =  \frac{J\wedge J\wedge J}{D_{ABC}J^A J^B J^C}\) as the basis of \(H^0(X;\zz)\) and \(H^6(X;\zz)\)\footnote{Our triple intersection numbers \(D_{ABC}\) are defined by \(\int_X \a_A \wedge\a_B \wedge\a_C =D_{ABC} \). For the readability of the equations, we will extensively avoid writing out all the \(D_{ABC}\)'s explicitly. It should therefore be understood that, for a vector \(k^A\) in \(H^2(X;\rr)\), \((k^2)_A \equiv D_{ABC} k^B k^C\) and \(k^3 \equiv D_{ABC} k^A k^B k^C\) etc. } .

In this basis we can write a general function with value in \(H^{2*}(X;\rr)\) as 
\be
\D = \D^\L \a_\L +  \D_\L \b^\L\;\;\;;\;\;\;\L = 0, 1, ... , b_2(X)\;.
\ee
For example, the charge vector of a BPS state can be written as
\be
\G = p^0 + p^A \a_A + q_A \b^A + q_0 \b^0\;.
\ee

In this notation, the symplectic product of two vectors in \(H^{2*}(X;\rr)\), which in the mirror picture is just the  symplectic product of two vectors in \(H^{3}(\til{X};\rr)\), reads
\be
<\D,\D'> = \int_X \D\wedge \D'^*\;\;\;;\;\;\;   \D^*\equiv \D^0 - \D^A \a_A + \D_A \b^A - \D_0 \b^0\;. 
\ee

To specify a four- and in turn a five-dimensional solution, we have to specify the background moduli \(\lim_{|\vec{x}|\rightarrow \inf} (B(\vec{x})+i J(\vec{x}) ) \) as well. We encode this information by defining \(\O(\vec{x}): \rr^3 \rightarrow H^{2*}(X;\rr)\) as 

\be
\O_{0} = - e^{(B+iJ)}\;\;;\;\;\; \O = \frac{\O_{0} }{\sqrt{i<\O_{0}, \bar{\O}_{0}>}} = - \frac{e^{(B+iJ)}}{\sqrt{\frac{4}{3} J^3}}\;,
\ee
\footnote{In this paper we will be working in the large charge regime and we will systematically ignore all the higher order corrections.}where \(B\), \(J\) are two-forms and it should be clear that by the exponential we really mean the terms in its expansion until the cubic term. The central charge of a state with charge vector \(\G\) and with given complexified K\"ahler moduli is then given by \(Z(\G;B+iJ) = <\G,\O>\).

The way that the internal moduli manifest themselves in the non-compact space is through the (2\(b_2\)+2) harmonic functions which completely determine the 4d and 5d metric: for a BPS state in type IIA on X with given background moduli \(\lim_{|\vec{x}|\rightarrow \inf}\O = \O|_\inf\) and with N centers with charge vectors \(\G_i\) and (pointlike) locations \(\vec{x}_i\) in \(\rr^3\), \(i =1 , ..., N\), the harmonic functions \(\Hf(\vec{x}): \rr^3 \rightarrow H^{2*}(X;\rr)\) are given by 

\bea \label{harmonics4}
\Hf &=& \Hf^\L \a_\L + \Hf_\L \b^\L = \sum_{i=1}^N \frac{\G_i}{|\vec{\xf}-\vec{\xf}_i|} + \hf \\
\hf& =& \hf^\L \a_\L + \hf_\L \b^\L= - 2 {\mathrm{Im}}\Bigl((e^{-i\th}\O)|_\inf\Bigr)\;,
\eea
where \(\th\) is the phase of the total central charge , \(Z(\G=\sum_i\G_i)=e^{i\th}  |Z(\G)| \).

Now we are ready to forget about the compact dimensions and focus on the non-compact ones. The metric part of the four- and five-dimensional solutions, of the low energy supergravity theories obtained by compactifying  IIA and M-theory on \(X\), are given respectively by

\be \label{metric4d}
ds_{4d}^2 = -\frac{\p}{\Sf} (d\tf + \o_{(4)})^2 + \frac{\Sf}{\p} d\xf^a d\xf^a\;\;\;\;a= 1,2,3\ee
and
\bea 
\label{lift_metric}
ds_{5d}^2 &=& 2^{2/3} ({\cal{V}}^{(s)})^2 (d\psi +A^0_{4D})^2 + 2^{-1/3}({\cal{V}}^{(s)})^{-1} ds_{4d}^2 \\ \nonumber
&=& 
-  (2^{2/3}\Qf)^{-2} ( \, ( d\tf + \o_{(4)} +2\Lf (d\psi + \o_{(4)}^0) )^2\\ \label{metric1}
&& + (2^{2/3}\Qf) 
\lbrace \frac{1}{\Hf^0}  (d\psi+\o_{(4)}^0)^2+   \Hf^0 d\xf^a d\xf^a \rbrace\;.
\eea \footnote{Here we use the special font and the subscripts ``(4)" to denote that these are the coordinates and functions natural from the four-dimensional point of view and are especially not suitable in the decompactification limit we will take, in which only the five-dimensional picture is valid. They have to be rescaled when we take the M-theory limit, as will be explained later in section 3.2.}

The 4d and 5d warp factors \(\Sf(\vec{x})\), \(\Qf(\vec{x})\) and the 5d rotation parameter \(\Lf(\vec{x})\) appearing here are functions of the \(\rr^3\) coordinates \(\xf^a\) and are given by the above harmonic functions  as 

\bea\label{entropyfunction}
\Sf &=&  2\p \sqrt{\Hf^0 \Qf^3 -  (\Hf^0\Lf)^2}\\
\Lf&=& \frac{\Hf_{0}}{2} - \frac{\Hf^A \Hf_{A}}{2\Hf^0} + \frac{D_{ABC}\Hf^A \Hf^B \Hf^C}{6(\Hf^0)^2}\\
\Qf^3 &=& (\frac{1}{6}D_{ABC}\yf^A \yf^B \yf^C)^2\\ 
D_{ABC}\yf^B \yf^C &=& -2\Hf_{A} + \frac{D_{ABC}\Hf^B \Hf^C}{\Hf^0}\;.
\eea

The cross terms in the 5d metric are determined up to coordinate redefinition by
\bea
d\o_{(4)} &=&\star^3_{(4)}<d\Hf,\Hf>\\
d\o^0_{(4)} &= & \star^3_{(4)} d\Hf^0\\ \nonumber
\mbox{   where the}&\star^3_{(4)}&\mbox{is the Hodge dual operator w.r.t. the flat } \rr^3 \;,
\eea
and the Calabi-Yau volume in string units \( ({\cal{V}}^{(s)})^3\) is given by
\bea
({\cal{V}}(\vec{x})^{(s)})^3 &=& \frac{1}{6}D_{ABC} \im t^A \im t^B \im t^C = (\frac{\Sf}{2\p} \frac{1}{\Hf^0 \Qf})^3\\
t^A &=&\frac{H^A - \frac{i}{\p} \frac{\pa S}{\pa H_A} }{H^0 + \frac{i}{\p} \frac{\pa S}{\pa H_0}}
\;.
\eea

To avoid repetition we leave the complete expressions for the vector multiplets part of the solution for section 3.2.

Furthermore, in order for \(\o_{(4)}\) to have a global solution, one has to impose an integrability condition \(d\,d\o_{(4)}=0\) which reads
\be \label{integrability}
<\Hf, \G_i>|_{\vec{\xf}=\vec{\xf}_i} =  0 \;\;\; \mbox{for all}\, i = 1,...,N\;.
\ee

This condition constrains the distances between the centers but is in general not sufficient to fix them once there are more than two centers. 

The four-dimensional warp factor \(\Sf(\vec{x})\) is also called the entropy function, which in the case of a single black hole indeed approaches \(S\rightarrow \frac{S_{bh}}{r^2}= \frac{A_{hor}}{4 (l_P^{(4)})^2} \frac{1}{r^2}\) when approaching the black hole horizon. In the general multi-center cases, on the other hand, it's not obvious that \(\Sf^2\) given in (\ref{entropyfunction}) is positive everywhere in the base space \(\rr^3\). From the four-dimensional point of view it is clear that the condition
\be
\Hf^0 \Qf^3 -  (\Hf^0\Lf)^2  \geq  0
\ee
 has to be satisfied in order to get a metric that is real everywhere. As we will show later, in the five-dimensional picture this condition manifests itself as the condition of the absence of closed timelike curves.  

\section{Construct the Bubbling Solutions}
\setcounter{equation}{0}

After reviewing the formulae we need, now we can construct the bubbling solutions in three steps: first taking the limit, second rescaling the solution, and finally specifying the centers.

\subsection{M-theory Limit}

First of all, in order to get an asymptotically flat metric in 5d, it is clear that one should take the decompactification limit in which the M-theory radius \(R_M\) goes to infinity. Recall the relation \(R_M = \ell_s g_s\) ,  \(\ell_P^{(11)} = \ell_s g_s^{1/3}\), from \(J^{(s)} \ell_s^2 = J^{(M)} (\ell_P^{(11)})^2\) one gets
\be
J^{(s)} = J^{(M)} \frac{R_M}{\ell_P^{(11)}} =  J^{(M)} (\frac{R_M}{\ell_s} )^{2/3}\;.
\ee

Therefore, if we require the Calabi-Yau volume to be finite in the eleven-dimensional Planck units, which is the criterion for the five-dimensional description to make sense, then taking \(\frac{R_M}{\ell_P^{(11)}} \rightarrow \inf\) in equivalent to taking the type IIA Calabi-Yau K\"ahler moduli \(J^{(s)} \rightarrow \inf\). 

We now therefore stipulate the background moduli to be 
\bea
B^A|_{\inf} &\equiv& b^A\mbox{    finite} \\
J^{A(s)}|_{\inf} &\equiv & j^A \rightarrow \inf\;.
\eea

In this limit the constant terms \(\hf\) in the harmonic functions take a especially simple form (the general expressions can be found in Appendix B):
\bea \label{hf1}
\hf^0\mbox{ ,  }\hf^A &\rightarrow& 0\\ \label{hf2}
\hf_A &\rightarrow& - \frac{p^0}{|p^0|}\, \frac{(j^2)_A}{\sqrt{\frac{4}{3}\,j^3}}\\ \label{hf3}
\hf_0 &\rightarrow& -\frac{1}{|p^0|}\frac{D_{ABC}p^A j^B j^C}{\sqrt{\frac{4}{3}\,j^3}} = \frac{p^A}{p^0} \hf_A \;.
\eea

\subsection{Rescale the Solution}

It seems that we are done with the background moduli and all still left to be done is to choose the appropriate charges and fill them in the harmonic functions. But there is a subtlety which is a consequence of the large (IIA) Calabi-Yau volume limit that we are taking. One can see this already from the expression for the constant terms in the harmonic functions (\ref{hf2}), (\ref{hf3}): these remaining constants go to infinity in this limit! Indeed, as a result, the three-dimensional (apart from the time and the 5th dimension) part of the metric goes to \((H^0Q)|_{\inf} d\xf^a d\xf^a \rightarrow \inf \,\frac{d\xf^a d\xf^a}{|\vec{x}|}\) at spatial infinity, while it goes to zero in the timelike direction: \( - g_{tt} = 2^{-4/3} \frac{1}{Q^2} \rightarrow 0\).\footnote{See the next section for detailed asymptotic analysis.} This is a clear signal that we are using a set of coordinates not appropriate in the five-dimensional description.

To find the right coordinates, let's remind ourselves that the four-dimensional metric is measured in the four-dimensional Planck units, while the extra warp factor \({\cal{V}}^{-1}\) rescale the metric to be measured in the five-dimensional Planck length when the the solution gets lifted (see (\ref{lift_metric}) ). The problem is really that, in the limit we are working in, the ratio between the five-dimensional Planck length \(\ell_P^{(5)}\sim \frac{\ell_P^{(11)}}{(\frac{1}{6}(J^{(M)})^3)^{1/3}}\) and the four-dimensional one \(\ell_P^{(4)} \sim \frac{\ell_s g_s}{(\frac{1}{6}(J^{(s)})^3)^{1/2}}\):  \(\frac{\ell^{(5)}_P}{\ell^{(4)}_P}\sim (\frac{(J^{(s)})^3}{6})^{1/6} \) goes to infinity. Therefore, in order to obtain a coordinate system natural in five dimensions, we should rescale all the coordinates with a factor \(\sim (\frac{(J^{(s)})^3}{6})^{1/6} \) and accordingly  the harmonic functions as well. Let's define

\bea 
\a&\equiv& \frac{1}{2} \, (\frac{4}{3}j^3)^{1/6} \\ \label{bigdef}
x^a&\equiv& \a \xf^a\\
t&\equiv& \frac{1}{2\a} \tf \\
\{H, L, Q , \o\}&\equiv&\frac{1}{\a} \{\Hf, \Lf, \Qf , \o_{(4)}\}\\
S &\equiv& \frac{1}{\a^2} \Sf
\eea

One can easily check that the lifted five-dimensional metric (\ref{metric1}) can be written in the above rescaled coordinates and functions in exactly the same form:
\bea \nonumber \label{metric2}
2^{-2/3} ds_{5d}^2 &=& - Q^{-2} \, \lbrack dt + \frac{\o}{2} + L (d\psi + \o^0) \rbrack^2 \\ && + Q 
\lbrack \frac{1}{H^0} (d\psi+\o^0)^2+   H^0 dx^a dx^a\rbrack \;.
\eea
The only difference the rescaling makes to the metric is that the warp factor \(Q(\vec{x})\) approaches a finite constant (\(= \pm\)1) even in the decompactification limit we are working in. 

Let's now pause and summarise. What we have done so far is to obtain a large number  of BPS solutions of five-dimensional supergravity with n vector multiplets, by lifting the four-dimensional solutions in the limit that the extra direction is infinitely large. These solutions might have singularities or/and horizons, depending on the charges of each center and their respective locations. For later use, we will now spell out explicitly the five-dimensional solutions. 
 
The metric part of the solution is given by (\ref{metric2}) and 

\bea \label{aflow1}
Q^3 &=& (\frac{1}{6}D_{ABC}y^A y^B y^C)^2\\ \label{flow2}
L&=& \frac{H_0}{2} - \frac{H^A H_A}{2H^0} + \frac{D_{ABC}H^A H^B H^C}{6(H^0)^2}\\
 \label{aflow2}
D_{ABC}y^B y^C &=& -2H_{A} + \frac{D_{ABC}H^B H^C}{H^0}\\ \label{o0}
\star^3 d\o &=& <dH,H>\\ \label{o}
d\o^0 &= & \star^3 dH^0\\ \nonumber
\mbox{   where the}&\star^3&\mbox{is the Hodge dual operator w.r.t.} \rr^3 \mbox{given by }x^a\;,
\eea
and the entropy function is again defined as
\be
S = 2\p \sqrt{H^0 Q^3 -  (H^0L)^2}\;.
\ee

The harmonic functions are given by, in their most explicit form:

\bea \label{5dharmonics1}
H^0(\vec{x})&=&\sum_i \frac{p^0_i}{r_{i}}\\\
H^A(\vec{x})&=& \sum_i \frac{p^A_i}{r_i}\\  \label{H_A}\\
H_A(\vec{x})&=&\sum_i \frac{q_{A,i}}{r_{i}}  +h_A\;\;;\;\;h_A=- \frac{|p^0|}{p^0}\frac{2\,(j^2)_A}{(\frac{4}{3}j^3)^{2/3}}\\  \label{5dharmonics4}
H_0(\vec{x})&=& \sum_i \frac{q_{0,i}}{r_{i}}  
+h_0\;\;;\;\;h_0=\frac{p^A}{p^0}h_A = -\frac{2}{|p^0|} \frac{D_{ABC}p^A j^B j^C}{(\frac{4}{3}j^3)^{2/3}}
\\ \nonumber 
\mbox{where } &r_{i}& = |\vec{x}-\vec{x}_i| .
\eea
Notice that now the remaining constant terms \(h_A\), \(h_0\) are insensitive to the rescaling of \(j\). We can therefore as well interpret the \(j\) as the M-theory asymptotic K\"ahler moduli \(j^A=\lim_{|\vec{x}|\rightarrow \inf}J^{A(M)}(\vec{x})\), which we keep as finite.

Since the integrability condition (\ref{integrability}) is going to play an important role in the analysis in the following section, we also rewrite it as
\be \label{integrability2}
<\G_i,H_i> = 0 \Leftrightarrow \sum_{j}\frac{<\G_i,\G_j>}{r_{ij}} = -h_A \til{p}^A_i \;,
\ee
where
\bea
H_{i} &\equiv& (H-\frac{\G_i}{r_i})|_{\vec{x}=\vec{x}_i}\\
\til{p}^A_i &\equiv& p^A_i - p^0_i \frac{p^A}{p^0} \;\;;\;\; 
r_{ij} = |\vec{x}_i-\vec{x}_j|\;.
\eea
Notice that the right hand side of (\ref{integrability2}) would in general have a much more complicated dependence on the charges of the centers, if we hadn't taken the M-theory limit.

Now we turn to the vector multiplets. The 4d vector multiplets are given by\footnote{Notice that convention for \( A^A_{4D}\) differs in sign from that of \cite{Bates:2003vx}; specifically, the coupling of a D0-D2 bound state with the gauge field is \(q_0 A_{4D}^0 + q_AA_{4D}^A \) in our convention.}
\bea
t^A &=&\frac{H^A - \frac{i}{\p} \frac{\pa S}{\pa H_A} }{H^0 + \frac{i}{\p} \frac{\pa S}{\pa H_0}}
=(\frac{H^A}{H^0} -\frac{L}{Q^{3/2}} y^A ) + i (\frac{S}{2\p} \frac{y^A}{ H^0 Q^{3/2}}) \\ 
A^0_{4D} &=& \frac{2}{S}\frac{\pa S }{\pa H_0} (dt+\frac{\o}{2})+ \o^0\\
 A^A_{4D} &=& \frac{2}{S}\frac{\pa S }{\pa H_A} (dt+\frac{\o}{2})- {\cal{A}}_d^A\;\;\;;\;\;\;d{\cal{A}}_d^A = \star_3 dH^A  \,. 
\eea

The lifted five-dimensional ones are given in terms of them as
\be
Y^A = \frac{\im t^A}{(\frac{(\im t)^3}{6})^{1/3} } = \frac{y^A}{(\frac{y^3}{6})^{1/3}},
\ee
satisfying \(\frac{Y^3}{6} = 1\), and
\bea \nonumber
A^A_{5D} &=& ( \re\; t^A ) (d\psi + A_{4D}^0) + A^A_{4D}\\ \label{A_5D}
&=& -\frac{y^A}{Q^{3/2}} (dt +\frac{\o}{2}) +(\frac{H^A}{H^0} -\frac{L}{Q^{3/2}} y^A )(d\psi+\o^0) - {\cal{A}}_d^A 
\;. 
\eea

In a form more familiar in the five-dimensional supergravity literature, these solutions can be equivalently written as
\bea
2^{-2/3} ds_{5D}^2 &=&  -Q^{-2}\,e^0 \otimes e^0 + Q ds_{base}^2\\ \label{field_strength_5d}
F^A_{5D} &=& d A^A_{5D} =  - d(Q^{-1}Y^A e^0 )  + \Theta^A \;,
\eea
where
\bea
ds_{base}^2 &=& H^0 dx^a dx^a + \frac{1}{H^0} (d\psi + \o^0)^2 \\
e^0 &=&  dt + \frac{\o}{2}  + L (d\psi + \o^0) \\
\Theta^A &=& \star_{base} \Theta^A= d\lbrack \frac{H^A}{H^0} (d\psi + \o^0) \rbrack - \star_3 dH^A \;.
\eea

\subsection{Specify the 4D Charges}
Now we would like to know what kind of 4d charges for the centers we should take, in order to obtain an asymptotically flat, smooth, horizonless solution when lifted to five dimensions. We now argue that the only possibility is the multi-center configurations composed of D6 and anti-D6 branes with world-volume fluxes turned on, and with the constraint that the total D6 brane charge equals to \(\pm1\).\footnote{Furthermore, each center must have D6 charge \(\pm 1\), if one also wants to exclude orbifold singularities at the center. But we will keep the formulae as general as possible and do not specify the D6 charges of each center.}
This can be understood as the following: if we take D2 or D4 branes or their bound states with other branes, the uplift to M-theory will have also M2, M5 brane sources and thus won't have the desired smooth and horizonless virtue. In other words, the uplifted metric near a D2 or D4 center will not be flat. One might also wonder about the possibility of adding D0 branes into the picture. First of all, in contrast to the usual scenario \cite{Witten:2000mf},  a D0-D6 bound state doesn't exist in the large volume \(J^{(s)}|_\inf \rightarrow \inf\) limit we are taking, irrespective of the (finite) value of the background B-field. But one could still imagine a multi-center KK monopole-electron-antimonopole-positron juxtaposition living in the large coupling limit. But this time the metric near the D0 centers is not smooth; more specifically, the metric in the 5th direction blows up while remaining flat in the \(\rr^3\) direction. In summary, in order to get a smooth and horizonless solution, we have to restrict our attention to D6 and anti-D6 branes with world-volume fluxes.

From the part of the D6 world-volume action coupling to the RR-potential \cite{Douglas:1995bn,Li:1995pq}
\be\label{wv_action}
\int_{\S_7} e^{B+F} \wedge C \;\;\;;\;\; C\in H^{2*}(X,\rr)\;,
\ee
one sees that the world-volume flux induces a D4-D2-D0 charge. Specifically, neglecting the B-field which can always be gauged into world-volume fluxes locally on the six brane, the charge vector of a center of \(p^0_i\) D6 and with world-volume two-form flux \(\frac{f_i}{p^0_i} =\frac{f_i^A}{p^0_i}  \a_A\) turned on is
\be \label{chargei}
\G_i = p^0_i e^{\frac{f_i }{ p^0_i }} = p^0_i  + f_i + \frac{1}{2} \frac{f_i^2}{p^0_i} +\frac{1}{6} \frac{f_i^3}{(p^0_i)^2}\;.
\ee

Thus the total charge vector is\footnote{In the case of stacked D6 branes, we only turn on the Abelian  fluxes. The reason for this restriction is that for non-Abelian \(F\), the induced D4-D2-D0 charges are proportional to \({\rm Tr} F\), \({\rm Tr} F\wedge F\) and \({\rm Tr} F\wedge F\wedge F\) respectively. In this case one can easily see that the corresponding solution will in general develop a singularity or a horizon. }
\bea \nonumber
\G&=&p^0 + p^A \a_A + q_A \b^A + q_0 \b^0 \\
\label{totalcharges}
&=& \sum_{i=1}^N \G_i =\sum_{i=1}^N p^0_i  + \sum_{i=1}^N f_i +\sum_{i=1}^N \frac{1}{2} \frac{f_i^2}{p^0_i} +\sum_{i=1}^N\frac{1}{6} \frac{f_i^3}{(p^0_i)^2}\;.
\eea
As mentioned earlier, we are especially interested in the case \(p^0 = \pm  1\), since this condition ensures asymptotic flatness. More specifically, only for the case \(p^0 = \pm  1\) the metric approaches that of \(\rr_t\times \rr^4\) in spatial infinity without identification.

Simply filling these charges into the harmonic functions in the last subsection gives us, as we will verify later, a metric that is asymptotically flat, smooth and horizonless everywhere, and may or may not have the conserved charges of those of a classical black hole. 

\section{The Properties of the Solution}
\setcounter{equation}{0}

\subsection{The Conserved Charges}

\subsubsection{4d and 5d Charges}

When lifting a four-dimensional solution to five dimensions, the charged objects in IIA get mapped into charged objects in M-theory. The Kaluza-Klein monopoles and  electrons, namely the D6 and D0 charges, show themselves as Taub-NUT centers and the angular momentum in the five-dimensional solution. Especially we expect \(q_0 \sim J_L\). The (induced) D4 charges, as can be seen in (\ref{field_strength_5d}), parametrize the magnitude of the part of the field strength that is self-dual in the Gibbons-Hawking base. In the type IIA language, in the case with non-zero D4 charges, one also has non-zero B-field in various regions in space. When lifted to M-theory they give a new contribution to the vector potential and we expect those to modify the definition of the electric charges. Therefore, as suggested in \cite{Gaiotto:2005gf}, \(q_{A,(5D)}\) and \(J_L\) will get extra contributions involving \(p^A\) through the Chern-Simons coupling and the Poynting vectors of the gauge field. An inspection of the five-dimensional attractor equation for a 5d black hole
\bea
S_{5D}&=& 2\p \sqrt{Q^3 - J_L^2} \\
Q^3 &=& (\frac{y_{(5D)}^3}{6})^2 \;\;;\;\;D_{ABC} y_{(5D)}^B y_{(5D)}^C = -2 q_{A,(5D)}\;,
\eea
and comparing it to the four-dimensional ones (\ref{aflow1}) and (\ref{aflow2}) with \(p^0=1\) suggests that, when \(p^A\) becomes non-zero, \(q_{A,(5D)}\) and \(J_L\) must get an extra contribution as
\bea
-2 q_{A,(5D)} &\rightarrow& -2 q_{A,(5D)}  + \frac{(p^2)_A}{p^0}
\\
J_L &\rightarrow&J_L- \frac{p^A q_A}{2p^0 } +\frac{p^3}{6\,(p^0)^2}\;.
\eea

We will now verify this through explicit asymptotic analysis, while more discussion related to the role of \(p^A\) charges can be found in section 4.3. 

\subsubsection{The Asymptotic Analysis}

Now we would like to work out the asymptotic form of the solution. We are interested in it for the following two reasons. First of all we would like to verify that our metric is indeed asymptotically flat; secondly we would like to read off all the conserved charges of these solutions.
The following asymptotic analysis applies to all the solutions in the form of that presented in the end of the last section, i.e., to {\it{all}} the solutions of the \(\cn=1\) five-dimensional supergravity obtained by lifting four-dimensional solutions in the decompactification limit. \footnote{Apart from the fact that we are assuming in this subsection that the sign of the total D6 charge is positive, to avoid messy phase factors everywhere. The adaptation to the case in which \(p^0 < 0 \) is straightforward. }

Let's first look at the metric part. In the limit \( r = |\vec{x}| \rightarrow \inf\) we have the various quantities in the metric approaching\footnote{One has to be a bit careful with the order of taking the two limits \( r \rightarrow \inf\) and \(j^A\rightarrow \inf\). Here we restrict ourselves to the range \( 1 \ll r \ll \frac{R_M}{\ell^{(5)}_P} \rightarrow \inf\), in other words, where the spacetimes remains appearing to be five-dimensional. In this range one can indeed ignore the extra constant terms \(h^0\), \(h^A\) (see Appendix B).  }
\bea
Q &=& 1 + \co(r^{-1})\\
H^0 &=& \frac{p^0}{r} +  \co(r^{-2})\\
\o^0  &=& p^0\cos\th d\f + \co(r^{-1}) \\ \nonumber
L &=& \frac{1}{r}\, \lbrack\,  (\frac{q_0}{2} - \frac{p^A q_A}{2p^0} + \frac{D_{ABC}p^A p^B p^C}{6(p^0)^2})  + \frac{\hat{r}}{p^0} \cdot (\sum_{i,j=1}^N \frac{ <\G_i,\G_j>}{4} \frac{(\vec{x}_i-\vec{x}_j) }{|\vec{x}_i-\vec{x}_j|} )  \rbrack\\&& + \co(r^{-2})\;,
\eea
where the second term in the last equation is derived from the dipole term in the expansion and we have used the integrability condition (\ref{integrability2}) to put it in this form. 

We have now a natural choice of coordinates of the \(\rr^3\) factor of the metric. This is because the dipole term picks out a unique direction in the spatial infinity. Let's now choose the spherical coordinate in such a way that the vector
\be
\vec{J}_R = \sum_{i,j} \vec{J}_{ij}  =\sum_{i,j}  \frac{ <\G_i,\G_j>}{4} \frac{\vec{x}_i-\vec{x}_j}{|\vec{x}_i-\vec{x}_j|}
\ee
points at the north pole. The second term in \(L\) can then be written as \(\frac{1}{p^0}\vec{J}_R \cdot \hat{r} = \frac{1}{p^0} J_R \cos\th\).

Finally, solving the \(\o\) equation asymptotically gives us
\be
\frac{1}{2}\o =  \frac{1}{r}\,J_R\,\sin^2\th d\f +  \co(r^{-2})\;,
\ee
up to trivial coordinate transformations. 

After a change of coordinate \(r= \r^2/4\), the metric at infinity now reads
\bea \nonumber  2^{-2/3} ds_{5D}^2 &=&
-\lbrace dt + \frac{4}{\r^2} \lbrack p^0  J_L (\frac{1}{p^0}d\psi + \cos\th d\f) + J_R (d\f+ \frac{1}{p^0}\cos\th d\psi )  \rbrack + \co(\r^{-4}) \rbrace^2 
\\  \label{metric_asymp}
 &+& p^0 \lbrace d\r^2 + \frac{\r^2}{4} \lbrack d\th^2 + \sin^2\th d\f^2 + 
(\frac{1}{p^0}d\psi + \cos\th d\f)^2 \rbrack+ \co(\r^{-2})\rbrace\;,
\eea
with
\bea \label{J_L}
J_L & = &\frac{q_0}{2} - \frac{p^A q_A}{2p^0} + \frac{D_{ABC}p^A p^B p^C}{6(p^0)^2}\\
 \label{J_R}
J_R &=& |\sum_{i<j} \frac{ <\G_i,\G_j>}{2} \frac{\vec{x}_i-\vec{x}_j}{|\vec{x}_i-\vec{x}_j|} | \;
\eea
being the two angular momenta, corresponding to the \(U(1)_L\) exact isometry and the  \( U(1)_R\) asymptotic isometry, generated by \(\xi^3_L = \pa_{\psi}\) and \(\xi^3_R =\pa_\f\) respectively.

Indeed we see that, the metric approaches that of a flat space without identification when \(|p^0| =1\). In that case it can be more compactly written as
 \bea \nonumber
 2^{-2/3} ds_{5D}^2 &=& -\lbrack dt + \frac{4}{\r^2} (J_L \s_{3,L} + J_R \s_{3,R} ) \rbrack^2  \\
&+& ( d\r^2 + \frac{\r^2}{4}  (\s_{1,L}^2+ \s_{2,L}^2 + \s_{3,L}^2)  ) + ...\\ \nonumber
&=& -\lbrack dt + \frac{4}{\r^2} (J_L \s_{3,L} + J_R \s_{3,R} ) \rbrack^2 + ( d\r^2 + \frac{\r^2}{4}  (\s_{1,R}^2\\&+& \s_{2,R}^2 + \s_{3,R}^2)  ) + ...
 \eea
 where the \(\s\)'s are the usual \(SU(2)_L\) and \(SU(2)_R\) invariant one-forms of \(S^3\) (see for example \cite{Gauntlett:1995fu}).

After working out the angular momenta we now turn to the electric charges of the 5d solutions. 
The gauge field part of the action of \(\cn=1\) 5d supergravity is \cite{Gunaydin:1983bi}
\be
S_{gauge} = \frac{1}{16\p G^{(5)}}\,\int G_{AB} \, F^A\wedge \star_5 F^B -\frac{1}{6} D_{ABC} \, F^A\wedge F^B\wedge A^C\;,
\ee
where the metric of the vector multiplets coupling is given by
\be
G_{AB} = \frac{1}{2}\,\lbrace \frac{\pa}{\pa y^A}\frac{\pa}{\pa y^B} \,\log (\frac{y^3}{6}) \rbrace|_{\frac{y^3}{6}=1 } \;.
\ee
The conserved electric charges are then given by the Noether charge
\bea \label{}
q_{A(5D)} &\equiv& \frac{16\p G^{(5)}}{V_{S^3} } \int_{S^3_\inf} \frac{\pa S}{\pa F^A} 
= \frac{2}{V_{S^3} }\,\int_{S^3_\inf} G_{AB} \star_5 F^B - \frac{1}{6} D_{ABC}\,F^B\wedge A^C\;,
\eea
where \(V_{S^3}\) is just the volume of a unit 3-sphere.

We need to know the asymptotic behaviour of the vector potential and the field strength in order to compute the charges. They are given by
\bea
A^A_{5D} &=& \frac{p^A}{p^0}d\psi -\frac{j^A}{(\frac{1}{6}j^3)^{1/3}} dt + \co(\r^{-2})\; \; (
+ \mbox{gauge transformation})\\
F^A_{5D} &=& -d(\frac{y^A}{\frac{1}{6}y^3}) \wedge dt + \co(\r^{-2})d\s  + \co(\r^{-3}) d\r\wedge \s\;.
\eea

From these equations it is clear that the Chern-Simons term does not contribute to the charges, and from
\bea \nonumber
G_{AB} F^B_{5D} &=& -\frac{1}{4} d\Bigl( \frac{y^B}{y^3/6} \Bigr) \left\{ \frac{\pa}{\pa y^B}\Bigl(\frac{(y^2)_A}{y^3/6}\Bigr)\, \right\}|_{\frac{y^3}{6}=1} \wedge dt +... \\  \nonumber
&=&  -\frac{1}{4} d(y^2)_A \wedge dt + ... \\
&=& \frac{1}{2} (q_A - \frac{(p^2)_A}{2p^0})\,(\frac{\r}{2})^{-3} dt\wedge d\r ...\;.
\eea
we get after integration
\be
\label{q_5d}
q_{A(5D)} = q_A - \frac{(p^2)_A}{2p^0}\;.
\ee

This finishes our analysis of the conserved charges of our solutions. As mentioned earlier, the expressions for the charges and for the the asymptotic metric (\ref{metric_asymp}), (\ref{J_L}), (\ref{J_R}) and (\ref{q_5d}) apply to {\it{all}}  solutions lifted from four dimensions in the infinite radius limit, i.e., all the solutions presented in section 3.2. For the specific case we consider in the last section (let's focus on the case \(p^0 = +1\)), they are given simply by the D6 charge and the flux of each center as
\bea
q_{A(5D)} &=& q_A - \frac{(p^2)_A}{2p^0} = \sum_i \frac{(\til{f}_i^2)_A}{2 p^0_i}\\
J_L &=& \sum_i \frac{\til{f}_i^3}{6 (p^0_i)^2}\\ 
J_R &=&  |\frac{1}{4} \,\sum_{i,j=1}^N \,p^0_ip^0_j \frac{f_{ij}^3}{6} \frac{\vec{x}_i-\vec{x}_j}{|\vec{x}_i-\vec{x}_j|}\,|\,
\eea
where 
\bea \label{til_f}
\til{f}_i^A  &\equiv& f^A_i - p^0_i (\sum_j f^A_j) \\ \label{f_ij}
f_{ij}^A & \equiv& \frac{f_i^A}{p^0_i}- \frac{f_j^A}{p^0_j} =\frac{\til{f}_i^A}{p^0_i}- \frac{\til{f}_j^A}{p^0_j} \;.
\eea

As we will see later, \(\til{f}_i^A\) has the physical interpretation as the quantity invariant under the gauge transformation, and \(p^0_i p^0_j f_{ij}^A \) has the interpretation as the fluxes going through the \(ij\)th ``bubble". 

\subsection{The Shape of the Solution}

After analysing the solution at infinity, now we would like to know more about the metric part, i.e. the shape, of these solutions. First of all we would like to spell out the criterion that the metric is free of pathological closed timelike curves. Having black hole physics in mind, we would also like to see if the solution exhibits a throat (AdS-looking) behaviour in some region. These two parts of the analysis, unless otherwise stated, apply to general solutions presented in section 3.2. 

There is another region of special interest here. Namely, we would like to explicitly verify our claim that the metric, provided that the CTC-free condition is satisfied, is smooth and horizonless near each center. As discussed in section 3.3, this property only pertains to the special charges (D6 or anti-D6 with fluxes) that we have chosen. 

\subsubsection{Closed Timelike Curves}

Before jumping into the equations, let's first make a detour and look at the four-dimensional metric (\ref{metric4d}) we started with. Apart from the integrability condition (\ref{integrability2}), it's apparent that we 
also need to impose the condition 
\be \label{CTC}
(\frac{S(\vec{x})}{2\p})^2 = H^0 Q^3 - (H^0)^2 L^2 \geq 0\;,
\ee
in order to have an everywhere real metric in four dimensions. Indeed, in the case this is not satisfied, the volume of the internal Calabi-Yau goes through a zero and things stop making sense in all ten dimensions.  

A look at the 5d metric:  
\be
2^{-2/3} g_{\psi\psi} = (\frac{S(\vec{x})}{2\p})^2 (\frac{1}{H^0 Q})^2\;,
\ee
makes it clear that as long as the 4D metric is real everywhere, the lifted metric has its 5th direction always spacelike. Furthermore, from 
\be
(\frac{S(\vec{x})}{2\p})^2 = H^0 Q^3 - (H^0)^2 L^2 \geq 0 \Rightarrow H^0 Q\geq 0\;,
\ee
it also ensures that the warp factor in front of the \(\rr^3\) part of the metric is always positive, and therefore another danger for CTC is also automatically eliminated. In more details, this is because the harmonic functions are real by default, and it's really the \(Q\), or rather the \(y^A\), attractor flow equations that are not a priori endowed with a real solution. 

Now we can worry about the more subtle \(-Q^{-2} (\frac{\o}{2})^2\) part of the metric. Looking at the equation for \(\o\)
\be
d\o = \star_3 <dH,H>\;, 
\ee 
one sees that the danger zone is the region very close to a center, since it's the only place where \(dH\) and \(H\) blow up. But as we will see later, the integrability condition always guarantees that \(\o\) actually approaches zero at least as fast as the distance to the center under inspection. We can therefore believe that this term poses no threat. To sum up, what we find is
\be \label{CTC2}
\mbox{4d metric real  } \Leftrightarrow \mbox{  5d metric no CTC}\;.
\ee

Of course, mapping one problem to the other does not really solve anything. Indeed, at the moment the author does not know of any systematic way of checking this condition. Especially, the integrability condition, while often ensures the real (4d) metric condition (\ref{CTC}) to be satisfied near a center, is in general not sufficient to guarantee that it is satisfied everywhere.\footnote{In the four-dimensional context, a conjecture about the equivalence between the existence of a solution with an everywhere well-defined metric with given background and charges, and the existence of a split attractor flow connecting the asymptotic moduli and the attractor points of all the centers, has been proposed and studied in \cite{Denef:2001xn}, \cite{Denef:2000ar}, and \cite{Denef1}. If this conjecture is indeed true, it provides us a more systematic way of studying the existence of multi-centered solutions.} On the other hand, this is how it should be, since: given \(N\) centers, the naive moduli space of their locations grows like \( (\rr^3)^N \), the number of distances between them grows like \(N^2\), but the number of integrability condition grows only like \(N\). Given the possibility that one can always {\it{a priori}} add one more pair of centers with opposite charges while still keeping the total charge unaltered, it seems extremely unlikely to be able to obtain a reasonable moduli space for BPS states with a given total charge, if there are no rules of the game other than the integrability condition. 

We finish this subsection by noting that our discussion here about the closed timelike curves, especially the conclusion (\ref{CTC2}), applies to all 4D-5D lift solutions irrespective of the background moduli. That is, it applies even without taking the decompactification limit.

\subsubsection{The Throat Region}

In section 4.1 we have seen that, when we look at the asymptotic region: \be  h \gg \frac{1}{r} \gg \frac{r_{ij}}{r^2}\;,\ee the harmonic function can be expanded, in the order of decreasing magnitude, as
\be
H= h + \frac{\G}{r} + \mbox{dipole terms} +\mbox{quadrupole terms}  + ... \;,
\ee
where the non-vanishing constant terms \(h\) are of order one in our renormalization (see section 3.2).

If the (coordinate) distances \(r_{ij}\) of each pair of centers are all much smaller than one, namely \(r_{ij} \ll 1\;\forall\; i, j\), one can consider another region in which 
\be \label{throat_region}
\frac{1}{r} \gg h \;\; , \mbox{      } \frac{1}{r}  \gg \frac{r_{ij}}{r^2}\;.
\ee
 In other words, when the centers are very close to each other, one can zoom in a bit more from the asymptotic region so that  the constant terms become subdominant, while still not getting substantially closer to any of the centers than the others, and can still see the conglomeration of centers (the blob) as an entity without seeing the structure of distinct centers. 

In this region, the harmonic functions are expanded, again with descending importance, as
\be
H=  \frac{\G}{r} + \Bigl( h + \mbox{dipole terms} \Bigr) +\mbox{quadrupole terms}  + ... \;,
\ee
and attractor flow equation is given by
\be
D_{ABC}y^B y^C = \frac{1}{r} \,(-2q_A + \frac{(p^2)_A}{p^0} ) + ... \;.
\ee
Define \(y^A_{bh}\) to be the solution to the equation \((y_{bh}^2)_A =-2q_A + \frac{(p^2)_A}{p^0} \) and \(Q_{bh}^3 = (\frac{y_{bh}^3}{6})^2\), one arrives at
\be
Q = \frac{Q_{bh}}{r} + ...  .\ee

At the same time, 
\be
L  = \frac{1}{r}\,J_L+ ... = \frac{1}{r}\,(\frac{q_0}{2}-\frac{p\cdot q}{ 2 p^0} + \frac{p^3}{6 (p^0)^2 }) + .... \,.
\ee
Notice that, unlike in the asymptotic region, the dipole contribution to \(L\) is sub-leading because now 
\( \frac{1}{r} \gg h\). Again using the integrability condition to relate the dipole contribution of \(L\) to the magnitude of \(\o\), one sees that \(\o\) as well is of minor importance in this region.

Now the 5th dimension part of  the metric reads
\be \label{CTC_throat}
g_{\psi\psi}  = 2^{2/3} \, (\frac{Q}{H^0} - \frac{L^2}{Q^2}) = \frac{1}{(p^0)^2 Q_{bh}^2
} (\frac{S_{bh}}{2\pi})^2 + ...\;,
\ee
where \be S_{bh}  = 2\p \sqrt{p^0 Q_h^3 - (p^0)^2 J_L^2}  \ee is a constant equal to the (classical) black entropy with the charges corresponding to that of the total charges of our multi-center configuration. 

Putting everything together, we find that the metric in the region (\ref{throat_region}) looks like\footnote{For the readability we have imposed in the this equation that the total monopole charge \(p^0=1\). It's trivial to put back all the \(p^0\) factors, and the metric one obtains in the case of \(|p^0|\neq 1\) is that of an orbifolded BMPV near horizon geometry.}
\bea
2^{-2/3} ds_{5D}^2 &=&  -(\frac{r}{r_{bh}})^2 dt_{bh}^2 + (\frac{r_{bh}}{r})^2 dr^2 + 2 r \,(\frac{J_L}{r_{bh}^3})\, dt_{bh}\,\s_{3,L} 
\\  \label{near_horizon_1} &+& r_{bh}^2 \Bigl( \s_{1,L}^2 +\s_{2,L}^2+ \s_{3,L}^2 -(\frac{J_L^2}{r_{bh}^3})^2 \s_{3,L}^2 ) \Bigr)\;,
\eea
where \(r_{bh}\equiv \sqrt{Q_{bh}}\) and we have rescaled the time coordinate \(t_{bh} = \frac{t}{\sqrt{Q_{bh}}}\). 

One can now readily recognise this metric as the \(AdS_2\times S^3\) near horizon metric of a BMPV black hole\footnote{Or, more precisely, an identification of \(AdS_3\times S^3\) which leaves a cross term \(dt \,\s_{3,L}\) behind \cite{Alonso-Alberca:2002wr}. Also the \(S^3\) is squashed in such a way that its area again gives the black hole entropy.}  \cite{Kallosh:1996vy}. Therefore we can identify the region (\ref{throat_region}) as a sort of near horizon region of the multi-center BPS solution.  

So far it all seems very satisfactory: the 5D solutions obtained from lifting multi-center 4D solutions have a throat region which looks like the near horizon limit of a classical black hole with charge given by the total charge of the 4D centers via the prescription we give in section 4.1. But we should not forget that the analysis here depends on the existence of the region (\ref{throat_region}). Indeed, it's obvious that this region cannot exist for all choices of charges: when the total charge does not give a classical black hole, namely when \(S_{bh}^2 < 0\), the existence of this region together with (\ref{CTC_throat}) would imply the presence of a CTC, or equivalently, an imaginary metric in 4D, in this region. One thus conclude that the region (\ref{throat_region}) can only exist when the total charge of all the centers together corresponds to that of a black hole. This also justifies our notation \(y_{bh},Q_{bh},t_{bh},r_{bh}\). 

In other words, when the total charge doesn't give a black hole, at least one pair of the centers must be far away from each other:
\be
\exists\;i, j \mbox{    s.t.    } r_{ij} \sim h \mbox{   or   }  r_{ij} > h  \mbox{      if    }  S_{bh}^2 < 0\;.
\ee
 
This argument applies actually not only to multi-center solutions in the large volume limit with arbitrary charges, but also to those with {\it{arbitrary}} background moduli \(j, b\), with the only difference being that we have to include in general much more complicated constant terms in the harmonic functions (see Appendix B) to estimate the lower bound on the distances between the centers. Therefore we conclude that, for a choice of charges such that the total charge doesn't give a black hole, the centers cannot get arbitrarily close to each other, at least as long as we stay in the regime where the supergravity description is to be trusted \(\frac{R_M}{\ell^{(11)}_P } \gg 1\); \(J^{(M)} \gg 1 \Leftrightarrow\) \(g_s \gg 1\) ; \(J^{(s)} \gg 1\). What happens to these multi-center configurations with total charge of no black holes, when \(\frac{R_M}{\ell^{(11)}_P} = g_s\) is lowered beyond the supergravity regime is described in terms of microscopic D-brane quiver theory and the higgsing thereof in \cite{Denef:2002ru}. From the five dimensional point of view, it would be interesting to refine the result of \cite{Dijkgraaf:2006um} in a similar spirit. 
 
We finish our throat examination with two remarks. First of all, the reverse of what we just said is not always true:  when the total charge does correspond to that of a classical black hole, the centers don't have to sit very close to each other. We can also imagine them to be far apart  and still have a well-defined metric. For example, the centers can split themselves up into two blobs far away form each other, with each blob having its throat region and can therefore be coarse-grained as an AdS-fragmentation kind of scenario \cite{Dijkgraaf:2005bp},\cite{Maldacena:1998uz}. 
Furthermore, it should be clear that our analysis given above does not exclude the presence of any kind of throat other than the ``common throat" encompassing all the centers as we discussed here. Especially, when the total charge of a subset of the centers corresponds to the charge of a black hole, one might also expect the presence of a ``sub-throat" encompassing just the subset in question, given that the other centers are sufficiently far away. The most well-known example of this phenomenon is that of the black ring geometry, which can be seen as the uplift of a D6 and a D4-D2-D0 center in the M-theory limit\cite{Elvang:2005sa,Gaiotto:2005xt,Bena:2005ni}. In the case that the total charge corresponds to that of a D6-D4-D2-D0 black hole (the case of small D0 charge), one has indeed a common throat of the BMPV type we discussed above. But apart from that, if one zooms in further near the D4-D2-D0 center there is another \(AdS_3\times S^2\) ``sub-throat" region, which is locally the same as the uplift of the D4-D2-D0 near horizon geometry and which gives the Bekenstein-Hawking entropy of the black ring\footnote{which is the same as the entropy of the D4-D2-D0 blak hole.}. For the special case of \(T^6\) compactification, a related issue is discussed in the dual D5-D1-P language in \cite{Elvang:2004ds,Bena:2004tk}.

Finally, the presence of a throat region opens the possibility to learn more about the CFT states these solutions correspond to: by treating the throat region as an asymptotically AdS spacetime, we can employ the AdS/CFT dictionary to read off the relevant vevs of these proto-black holes, see for example \cite{Skenderis:2006ah}. It will be interesting to see what kind of CFT states our bubbling solutions (including the known ones of Bena-Warner {\it{et al}}) correspond to. 

\subsubsection{Near a Center}

While much of the discussion above applies generally to all the lifted solutions in the large radius limit and depend only on the total charges, the solution near a center is of course strongly dependent on how the charges are allocated. Indeed, as we discussed in section 3.3, we've chosen the specific D6 and anti-D6 with Abelian world-volume fluxes as our centers because we'd like the metric to be free from horizons and singularities. Now we will explicitly verify this by analysing the metric near a center. Therefore, unlike most of the equations in the previous subsections, our discussion here applies only to the charges we described in section 3.3:
\be
\G=
\sum_{i=1}^N \G_i =1 + \sum_{i=1}^N f_i +\sum_{i=1}^N \frac{1}{2} \frac{f_i^2}{p^0_i} +\sum_{i=1}^N\frac{1}{6} \frac{f_i^3}{(p^0_i)^2}\;.
\ee

In the region very close to the {\it{i}}th center, where
\be
\frac{1}{r_i} \gg \frac{1}{r_{ij}}, \;h_0,\;h_A\;,
\ee
we can expand the harmonic functions as
\be
H = \frac{\G_i}{r_i } + H_i +\co{(\frac{r_i}{r_{ij}^2})}\;,
\ee
with \(H_i\) defined below (\ref{integrability2}).

If we plug this into the attractor flow equation, and notice that the possible \(\frac{1}{r_i}\) term cancels because our choice of charges has the virtue
\be
-2q_{A,i} + \frac{(p_i)_A^2}{ p^0_i} = 0 \;,
\ee
we get
\be
D_{ABC}y^B y^C = -2 c_{A,i} + \co(\frac{r_i}{r_{ij}})\;,
\ee
where
\bea
c_{A,i} &=& H_{A,i}+\frac{1}{p^0_i}H^0_i \,q_{A,i} - \frac{1}{p^0_i} D_{ABC} p^B_i H^C_i \\
&=& h_A + \sum_j \frac{p^0_j}{r_{ij}}\frac{(f_{ij}^2)_A}{2}
\eea
is a constant. 

The condition that the \(\rr^3\) part of the base metric is positive \( QH^0 > 0 \) can be satisfied if \be 
\label{near_center_condition} p^0_i c_{A,i} < 0\,.\ee

Assuming that our choice of locations and fluxes satisfies this condition, we have a solution
\bea
y^A &=& y^A_i +  \co(\frac{r_i}{r_{ij}}) \mbox{   where}\\
\frac{(y_i^2)_A}{2}  &=& -c_{A,i}\\
\Rightarrow Q^3 &=& Q^3_i + \co(\frac{r_i}{r_{ij}}) = (\frac{y_i^3}{6})^2 + \co(\frac{r_i}{r_{ij}})\;.
\eea
 
With a similar expansion and exploit the integrability condition (\ref{integrability2}) at the \(i\)th center and the explicit expression of the charges (\ref{chargei}), we get  
 \bea
 L &=& \co(\frac{r_i}{r_{ij}}) \\
  \o^0 &=& p^0_i \cos\th d\f + \co(\frac{r_i}{r_{ij}})\\
d\o &=& \star_3<dH,H> = \star_3 dr_i \, \co(\frac{1}{r_{i}})\\
 \Rightarrow  \o &=& \co(r_i)\;.
 \eea

Notice here that the first equation guarantees that (\ref{near_center_condition}) is enough to ensure that there is no closed timelike curve near this center. 

With everything put together, we obtain the metric near the \(i\)th center:
\be
2^{-2/3} ds_{5D}^2 = -dt'^2 + d\r^2 +\frac{\r^2}{4} \lbrack d\th^2 + \sin^2\th d\f^2 + (\frac{1}{p^0_i}d\psi+\cos\th d\f)^2 \rbrack + + \co(\frac{r_i}{r_{ij}})\;,
\ee
where we have rescaled the coordinates as \(t' = \frac{t}{Q_i}\), \(\r^2 = 4p^0_i Q_i r_i\). Therefore we conclude that metric approaches that of a \(\cc^2/ \zz_{p^0_i}\) orbifold, and has nothing more singular than a usual orbifold singularity. Specifically, the solutions with only \(p^0_i = \pm 1\) for all the centers will be completely smooth everywhere.

Furthermore, one sees that the \(U(1)_L\) isometry generated by \(\xi_L^3 = \pa_\psi\) has a fixed point at the center. Thus a non-trivial two-cycle which is topologically a sphere (the bubbles) is formed between any two centers and therefore the name ``bubbling solutions" (or rather the ``sausage network" solutions).  These two-cycles can support fluxes and indeed, the fluxes going through the \(ij\)th bubble is \(p^0_i p^0_j {f_{ij}^A}\), with \(f_{ij}^A\) defined as (\ref{f_ij}) \cite{Berglund:2005vb}. Furthermore, the amount of fluxes going through the bubbles constrains the distance between them through the integrability condition (\ref{integrability2}), which in this case reads
\be
\sum_j \frac{1}{r_{ij}} \,p^0_i p^0_j \frac{f_{ij}^3}{6} = -h_A \til{p}^A_i = -h_A \til{f}^A_i \;. 
\ee

\subsection{Large gauge Transformation}

It is well known that there is a redundancy of description, namely a gauge symmetry, in type IIA string theory or equivalently M-theory, which is related to the large gauge transformation of the B-field and the three-form potential \(C^{(3)}\) respectively. Physically, this large gauge transformation can be incurred by the nucleation of a virtual M5-anti-M5 pair and thus the formation of a Dirac surface in five dimensions \cite{deBoer:2006vg}. This shift of \(C^{(3)}\) also shifts the definition of the charges, but leaves all the physical properties of the solution intact. 

While this is a generic feature for all choices of charge vectors and all background moduli one might begin with, what we are going to do here is just to check this gauge symmetry explicitly for our bubbling solutions. 

Indeed, in our case, the transformation
\be \label{gauge_sym}
f_i^A \rightarrow f_i^A + p^0_i c^A\;\;\;;\;\;\; c^A \in \zz^{b_2(X)}
\ee
will in general change the charges (\ref{chargei}) of the configuration, especially the total D4 charge will transform like
\be
p^A \rightarrow p^A + c^A
\ee
in the case \(p^0 =1\). 
Especially, one can always exploit this symmetry to put \(p^A=0\).
It's trivial to check that the quantities \(Q, L, \o, \o^0\) in the metric are also invariant under this transformation, since all the combinations of harmonic functions involved can equally be written in terms of the ``invariant flux parameters" \(\til{f}_i\) and \(f_{ij}\) defined in (\ref{til_f}) and (\ref{f_ij}). Especially, all the conserved charges are invariant under the transformation. On top of that, we see that the right hand side of the integrability condition (\ref{integrability2}) is also invariant.\footnote{In general, in the four-dimensional language, this also implies that the existence of a BPS bound state of given, fixed charges such that \(\til{p}^A_i= p^A_i -\frac{p^A}{p^0} p^0_i \neq 0\) for every center, is insensitive to the shift of B-field in the large volume limit.} We can therefore conclude that the metric part of the solution has a symmetry (\ref{gauge_sym}). 

Furthermore, a look at the gauge field (\ref{A_5D}) tells us that this transformation indeed corresponds to a large gauge transformation of the \(A_{5D}^A\); equivalently, in the full eleven and ten dimensions, it corresponds to 
\be
C^{(3)} \rightarrow C^{(3)}  + c^A d\psi \wedge \a_A \mbox{  (M-theory)}\;\;\;;\;\;\;B\rightarrow B + c^A \a_A \mbox{  (IIA)}\;. 
\ee

Indeed, a look at the D6 brane world-volume action (\ref{wv_action}) makes it clear that the transformation (\ref{gauge_sym}) can be seen as turning on an extra integral B-field. This explains the origin of this extra symmetry.

\section{Conclusions and Discussion}
\setcounter{equation}{0}

What we have done in this note is to motivate and present a large number of asymptotically flat, smooth, and horizonless solutions to the five-dimensional supergravity obtained from the Calabi-Yau compactification of M-theory. We also analysed their various properties and along the way described various properties of  generic five-dimensional solutions obtained from lifting the multi-center four-dimensional solutions. 

A natural question to ask is the degeneracies of such solutions. From our analysis it is obvious that these bubbling solutions we describe have the same degeneracies as their four-dimensional counterparts. Especially, these are charged particles without internal degrees of freedom; their degeneracies have to come from the non-compact spacetime. 

Relatively little is known about the degeneracies of such states, though. The core of this supergravity problem is really that, although we have the integrability condition (\ref{integrability2}) to constrain the type of the solutions we can have, generically it is not enough. Indeed, while in many cases this condition alone can exclude the existence of a bound state of given charges and background moduli, generically the fact that it can be satisfied does not mean that the solution has to exist.  
Another criterion a valid solution has to conform to is the real metric condition (\ref{CTC}), which gets translated in five dimensions as the no CTC condition. Though the integrability condition helps  to exclude the presence of an imaginary metric near a center, in general it does not guarantee anything. For the purpose of counting bubbling solutions and also for the greater ambition of counting multi-center degeneracies in general, it would be extremely useful to have a systematic way to see when the integrability is enough and when we have to impose additional conditions, and of what kind. 

For the case that is of special interest, that is the case in which the total charge is that of a black hole, the problem is also of special difficulty. The situation is described in \cite{Denef:2002ru} as the following: if we tune down the string coupling, at certain point the distances between the centers will be of the string length (recall that \(\ell^{(4)}_P \sim \frac{l_s g_s}{\sqrt{(\frac{(J^{(s)})^3}{6})}}\)) and the open string tachyons will force us to end up in a Higgs branch of the D-brane quiver theory and thus a wrapped D-brane at one point in the non-compact dimensions. But in the other direction, for the case with a black hole total charge at least, things are much more complicated. As one increases the \(g_s\), {\it{a priori}} the state doesn't necessarily have to open up, but rather it can just collapse into a single-centered black hole, or any other kind of possible charge splittings. Therefore, seen from this cartoon picture, the D-brane degeneracy really has to be the sum of degeneracies of all of the allowed charge splittings. While at the same time, if the total charge doesn't give a black hole, from the real metric condition (\ref{CTC}) we see that the system has to split up when \(g_s\) is tuned up, since these charges only have multi-centered configurations as supergravity embodiments. 

Now let's come back to the quest of smooth, horizonless solutions with black hole charges. We have argued that the bubbling solutions we presented seem to be the only kind of solutions which can be lifted from four dimensions with these virtues. In any case it would be interesting to find explicit BPS solutions to the 5D supergravity of M-theory on Calabi-Yau {\it{without}} any exact \(U(1)\) isometry. For example, some wiggly ring structure or other things our imagination permits. These can of course never be obtained by lifting 4D solutions. 

We will now finish this paper by some speculative comments on black hole entropy.
As we have mentioned in the introduction, the contrast between the conventional view on black holes and the one suggested by Mathur and collaborators is somewhat heightened in the setting of a general Calabi-Yau compactification. Let's first consider the proto-example of the fuzzball picture, in which one has a D1-D5 system on \(T^4\times S^1\), which can be related by a chain of dualities to an F1-P system. This system doesn't have classical entropy and the microscopic entropy \(S_{micro} = 2\p \sqrt{n_F n_P } \) comes from different modes of vibrations on the string. In this case, with some hindsight wisdom, one can argue that  it is not so surprising after all that one can actually construct the supergravity solutions describing the microstates \cite{Lunin:2001fv,Taylor:2005db}, since in this case the origin of the degeneracy {\it{is}} in the non-compact directions. Now we can just naively compare this with the case of a usual D4-D2-D0 Calabi-Yau black hole, whose entropy can be microscopically described by that of a MSW string \cite{Maldacena:1997de}. For a MSW string the microscopic entropy is given by \(S=2\pi \sqrt{\hat{q}_0(\frac{c_L}{6})}\), where \(\hat{q}_0\) plays the role of \(n_P\). On the other hand, the central charge \(c_L= D_{ABC}p^A p^B p^C\), with \(p^A\) being the M5 brane charges, has its in general by far the most important contribution from the degrees of freedom of deforming the M5 brane (the divisor) within the Calabi-Yau and the bundle on it. From this point of view it is puzzling to think about how in this case one could reproduce this degeneracy from the configurations in the non-compact directions. This is definitely a point that requires further understanding.

\section*{Acknowledgments}

I would like to thank Dieter Van den Bleeken for sharing his notes on the lift of other kinds of solutions, Frederik Denef and Kostas Skenderis for useful conversations, and Erik Verlinde for the initiative, lots of discussions and encouragement. 

This research is supported financially by the Foundation of Fundamental Research on Matter (FOM).

\appendix

\section{Reproduce the old Bubbling Solutions}
\setcounter{equation}{0}

The known bubbling solutions are given by (See \cite{Berglund:2005vb,Bena:2005va,Bena:2006is,Bena:2006kb})

\bea \nonumber
ds_{5d(b)}^2 &=& - (\frac{1}{Z_1 Z_2 Z_3})^{\frac{2}{3}}\,(dt+ k )^2 \\
&+& (Z_1 Z_2 Z_3)^{\frac{1}{3}} \lbrace \frac{1}{V} (d\psi + \O^0)^2 + V dx^a dx^a \rbrace \\
\eea

where

\bea
V&=& \sum_{i=1}^{N} \frac{p^0_i}{r_i}\;\;;\;\; r_i = |\vec{x} - \vec{x}_i |\;\;;\;\sum_{i=1}^N p^0_i = 1\\ 
L_A &=& 1- \frac{1}{2} D_{ABC} \sum_i \frac{1}{r_i } \frac{f_i^B f_i^C }{ p^0_i}\\
K^A &=& \sum_i \frac{f_i^A}{r_i } \\ \nonumber
M & =& -\frac{1}{2} \sum_i \sum_A  f_i^A \,+ \frac{1}{12} \sum_i  \frac{1}{r_i }  \frac{f_i^3 }{ (p^0_i)^2}\\
d\O^0  &=& \star^3 dV\\
k &=& \m (d\psi + \O^0) + \O\\
Z_A 
&=& L_A + \frac{1}{2V} D_{ABC} K^J K^K  \;\;\;;\;\;D_{ABC} = |\e_{ABC}| \\ 
\m &=& M + \frac{1}{2V} K^A L_A + \frac{1}{6V^2} K^3\\ \label{tilo}
\nabla \times \O &=& V \na M - M \na V + \frac{1}{2} (K^A \na L_A - L_A \na K^A)
\eea

Let's now see how our solutions contain these as a special case. 

Firstly, apply the formulae to the special 3-charge (STU) case
\be 
D_{ABC} = |\e_{ABC}| \;\;\;\;A, B, C = 1,2,3\;.
\ee

In general, the attractor flow equation (\ref{aflow1}) and (\ref{aflow2}) are difficult to solve, but not in this case:

\bea
Q^3 &=&  (\frac{1}{6}D_{ABC}y^A y^B y^C)^2 =  (y^1 y^2 y^3)^2\\
y^2 y^3 &=& -H_1 + \frac{H^2 H^3}{H^0}\mbox{      and permutations}\\ \label{Q1}
\Rightarrow Q ^3 &=&  (-H_1 + \frac{H^2 H^3}{H^0})
(-H_2 + \frac{H^1 H^3}{H^0})(-H_3 + \frac{H^1 H^2}{H^0})\;.
\eea

Secondly we take the special Ansatz that the K\"ahler form is the same in the asymptotics for all the three directions:
\be J^1|_\inf =J^2|_\inf =J^3|_\inf = j \rightarrow \inf \;,
 \ee
and that the background B-field is finite
\be B^A|_\inf = b^A \ll j  \;.
\ee

In this case we have 
\bea H_A & = &\frac{1}{2} \sum_i \frac{1}{r_{i}} \frac{(f_i)^2_A}{p^0_i} - 1\;\;\;\;A=1,2,3 \\
H_0&=& \frac{1}{2} \sum_i \frac{1}{r_{i}} \frac{(f_i)^3}{(p^0_i)^2} -  \sum_i (f_i^1+f_i^2+f_i^3)\;.
\eea

Now, if we rename the coordinates and quantities appearing in our solution as 
\bea
V&=& H^0\\
L_A &=& -H_A\\
K^A &=& H^A\\
M &=& \frac{H_0}{2} \\
\O &=& \frac{1}{2}\o\\
\O^0 &=& \o^0\\
\m &=& L \\
\Rightarrow Q^3 &=&  Z_1 Z_2 Z_3\;,
\eea
one can easily check that our solution (\ref{metric2}) reduces to 
\be ds_{5d}^2 = 2^{2/3} ds_{5d(b)}^2 \;,\ee and the equations for and relations between quantities defined in our solutions correctly reproduce those appearing in the known bubbling solutions.

\section{Constant Terms for General Charges and Background}
\setcounter{equation}{0}

\bea
Z&=&<\G,\O> =  \frac{1}{\sqrt{\frac{4}{3}J^3}}\,
\left(p^0\,\frac{(B+iJ)^3}{6}  - \frac{p\cdot(B+iJ)^2}{2}   + q\cdot(B+iJ) -q_0  \right)\\ \nonumber
\hf &=& -2 \im \Bigl( (e^{-i\th} \O )|_\inf \Bigr) \\ \nonumber
&=&\frac{2}{\sqrt{\frac{4}{3}j^3}}\,\frac{1}{|p^0\,\frac{(b+ij)^3}{6}  - \frac{p\cdot(b+ij)^2}{2}   + q\cdot(b+ij) -q_0|} \; \im\lbrace \\ \nonumber
&&\lbrack p^0\,\frac{(b-ij)^3}{6}  - \frac{p\cdot(b-ij)^2}{2}   + q\cdot(b-ij) -q_0  \rbrack \\ 
&&\cdot\lbrack \frac{(b+ij)^3}{6} + \frac{(b+ij)^2}{2} + (b+ij) + 1 \rbrack
\rbrace
\eea

\bea \nonumber
\hf^0&=& \frac{2}{\sqrt{\frac{4}{3}j^3}}\,\frac{1}{|p^0\,\frac{(b+ij)^3}{6}  - \frac{p\cdot(b+ij)^2}{2}   + q\cdot(b+ij) -q_0|}\\ && \lbrace \frac{p^0}{6} (j^3-3jb^2) + pjb - qj \rbrace \\ \nonumber
\hf^A &=& \frac{2}{\sqrt{\frac{4}{3}j^3}}\,\frac{1}{|p^0\,\frac{(b+ij)^3}{6}  - \frac{p\cdot(b+ij)^2}{2}   + q\cdot(b+ij) -q_0|}\\ && \lbrace b^A\, \lbrack \frac{p^0}{6} (j^3-3jb^2) + pjb - qj  \rbrack \\ \nonumber
&&+ j^A \,\lbrack  \frac{p^0}{6} (b^3-3j^2b) -\frac{p(b^2-j^2)}{2} + qb - q_0 \rbrack \rbrace
\\ \nonumber
\hf_A& =& \frac{2}{\sqrt{\frac{4}{3}j^3}}\,\frac{1}{|p^0\,\frac{(b+ij)^3}{6}  - \frac{p\cdot(b+ij)^2}{2}   + q\cdot(b+ij) -q_0|}\\ \nonumber && \lbrace \frac{(b^2-j^2)_A}{2}  \,\lbrack \frac{p^0}{6} (j^3-3jb^2) + pjb - qj  \rbrack \\ \nonumber
&&+ (jb)_A \,\lbrack  \frac{p^0}{6} (b^3-3j^2b) -\frac{p(b^2-j^2)}{2} + qb - q_0 \rbrack \rbrace
\\ \nonumber
\hf_0&=& \frac{2}{\sqrt{\frac{4}{3}j^3}}\,\frac{1}{|p^0\,\frac{(b+ij)^3}{6}  - \frac{p\cdot(b+ij)^2}{2}   + q\cdot(b+ij) -q_0|}\\ && \lbrace 
\frac{b^3-3j^2b}{6}\,\left(pjb-qj\right) \\ \nonumber
&& -\frac{j^3-3jb^2}{6}\,\left(  -\frac{p(b^2-j^2)}{2} + qb -q_0 \right)
\rbrace
\eea

\section{An Alternative Formulation}
\setcounter{equation}{0}

While the attractor flow equation and the 5-dimensional solution given in the subsection 3.2 is similar in form to those in the literature, we would like to present an alternative and equivalent formulation of them here.  
The motivation for doing this is the following: In equations (\ref{metric2}), (\ref{aflow1}) - (\ref{aflow2}) and (\ref{A_5D}), the harmonic function \(H^0\) seems to play a very special role. The solution seems to be hopelessly in peril when near the zero locus of \(H^0\): signature of the Gibbons-Hawking base space flips, various quantities in the metric and the gauge potential blow up; it is not at all obvious that the solution makes sense along the co-dimension one hypersurface  \(H^0=0\). In the context of the present paper we  are interested in the solutions with the property of being smooth everywhere, therefore we would have to check in particular that this also holds when \(H^0 \rightarrow 0\). This has indeed been done in a similar context in the previous work on bubbling solutions \cite{Berglund:2005vb,Bena:2005va} by explicitly checking that the divergences in various quantities cancel to high enough orders of \(H^0\) expansion. Instead of doing the same, we will present another way of writing the attractor flow equation and the 5-dimensional solution (\ref{metric2}), (\ref{aflow1}) - (\ref{aflow2}) and (\ref{A_5D}), such that it becomes manifest that there is no more danger near the hypersurface \(H^0=0\) than in other regions in the spacetime, for any choice of charges. 

Furthermore, slightly outside the context of the present paper, one sees that for a configuration with total D6 charge zero in the type IIA language, one has \(H^0\rightarrow 0\) in all directions in the asymptotically flat  region. To be able to deal with this class of multi-centered solutions, it is also useful to have a reformulation which naturally accommodates the zero locus of \(H^0\). 

Instead of writing the attractor flow equation in terms of \(y\) and \(Q\) as in (\ref{aflow1}) and (\ref{aflow2}), let's consider a function \(\i^A\) satisfying
\be
D_{ABC} (H^B+ H^0 \i^B) (H^C+ H^0 \i^C) = H^0 D_{ABC} y^B y^C\;.
\ee
After some algebra one arrives at the alternative formulation
\bea \nonumber
2^{-2/3} ds_{5d} &=& -(\frac{H^0}{q} )^2 (dt+\frac{\o}{2})^2 - 2 \frac{ \ell}{q^2}\, (dt+\frac{\o}{2})(d\psi+ \o^0)\\
&&+ \frac{\l}{q^2} \left(2\ell + (H^0)^2 \l\right) (d\psi+ \o^0)^2 +q\, dx^a dx^a\\ \nonumber
A^A_{5D} &=& -\frac{1}{q^{3/2}} \{ H^0(H^A + H^0 \i^A) (dt+\frac{\o}{2}) + (\ell \i^A - H^0 H^A \l) (d\psi+ \o^0)\} \\ && - {\cal{A}}_d^A \\
t^A&=& \frac{1}{q^{3/2}}\{ -\ell \i^A + H^0 H^A \l +\frac{i}{2\p} \sqrt{2\l\ell+ (H^0)^2 \l^2}\, (H^A+H^0\i^A) \}\;,
\eea
where \(\i^A\) (instead of \(y^A\)) satisfy
\be
D_{ABC} H^B \i^C = -H_A -\frac{H^0}{2} D_{ABC} \i^B \i^C
\ee
and \(\l\) and \(\ell\) (instead of \(Q\) and \(L\)) are defined as
\bea
\l&=& -\frac{H_A\,\i^A}{3} -\frac{D_{ABC}}{12} H^A \i^B \i^C - \frac{H_0}{2}\\
\ell&=& (H^0)^2 L = \frac{D_{ABC}}{6} H^A H^B H^C -H^0 \frac{H^A H_A}{2}  + (H^0)^2\frac{H_0}{2}
\eea
and \(q\) is a convenient shorthand for
\be
q^{3/2} = \ell + (H^0)^2 \l \;.
\ee

Furthermore, the real-metric/no CTC constraint (\ref{integrability2}) now  reads
\be
\l( 2\ell +  (H^0)^2 \l ) > 0\;.
\ee

Notice that now none of the quantities \(\i\), \(q\) or \(\ell\) nor any combination of them appearing in the solution diverges when \(H^0 \rightarrow 0\), therefore we have shown that the region where \(H^0\) vanishes is not more susceptible to singularity than any other generic one in the spacetime, and this holds irrespective of the charges.\footnote{Although it is true that the coordinate \(t\) might not be the most appropriate time coordinate near \(H^0=0\).} One can also see that \(\ell\) and \(\l\) instead of \(L\) and 
\(Q\) are indeed the natural functions to consider with physical relevance when \(H^0 \rightarrow 0\), by considering the case of a D4-D2-D0 black hole for example. After replacing the harmonic functions \(H\) by the charge vector \(\G\) one obtains\footnote{and after the appropriate translation between conventions: \(D_{ABC} \rightarrow \frac{1}{6}D_{ABC} \) and \(q_{0,A} \rightarrow - q_{0,A} \). } 
\bea
\ell&=& \frac{c_L}{6}\\
\l&=& \frac{1}{2} (-q_0 +\frac{1}{2} D^{AB}q_A q_B )  = \frac{1}{2} \hat{q}_0
\eea
as the quantities appearing in the microscopic description of the black hole entropy \cite{Maldacena:1997de}
\be
S= 2\p\sqrt {\frac{c_L}{6} \hat{q}_0} = 2\p\sqrt{2\l\ell}\;.
\ee

\end{document}